\documentclass[]{raa}
\usepackage{graphicx,times}
\usepackage{natbib}
\usepackage{amssymb,amsmath}
\bibpunct{(}{)}{;}{a}{}{,}

\usepackage[a4paper=true,dvipdfm=true,pagebackref=true]{hyperref}
\hypersetup{pdftitle = The title of my PDF, pdfauthor = My name, pdfsubject= The subject, pdfkeywords = keyword1 keyword2 keyword3}
\hypersetup{colorlinks = true, linkcolor = green, anchorcolor = red, citecolor = blue, filecolor = red, pagecolor = red, urlcolor = red}

\begin{document}

   \title{Heavy element contributions of rotating massive stars to Interstellar Medium
$^*$
\footnotetext{\small $*$ Supported by the National Natural Science Foundation of China.}
}

 \volnopage{ {\bf 20XX} Vol.\ {\bf X} No. {\bf XX}, 000--000}
   \setcounter{page}{1}

   \author{Ruiqing Wu\inst{1}, Chunhua Zhu\inst{1}, Guoliang L\"{u}\inst{1}, Zhaojun Wang
      \inst{1},  Helei Liu\inst{1}
   }

   \institute{School of Physical Science and Technology, Xinjiang University, Urumqi 830046,
China; {\it ruiqingwu163@163.com, chunhuazhu@sina.cn}\\
\vs \no
   {\small Received 2020 December 9; accepted 2021 January 12}
}

\abstract{Employing the the stellar evolution code (Modules for Experiments in Stellar Astrophysics),
we calculate yields of heavy elements from massive stars via stellar wind and
core$-$collapse supernovae (CCSN) ejecta to interstellar medium (ISM).
In our models, the initial masses ($M_{\rm ini}$) of massive stars are taken from 13 to 80 $M_\odot$,
their initial rotational velocities (V) are 0, 300 and 500 km s$^{-1}$, and their metallicities are
$[Fe/H]$ = -3, -2, -1, and 0.
The yields of heavy elements coming from stellar winds are mainly affected by the stellar rotation
which changes the chemical abundances of
stellar surfaces via chemically homogeneous evolution, and enhances mass-loss rate.
We estimate that the stellar wind can produce heavy element yields of about $10^{-2}$ (for low metallicity models)
to several $M_\odot$ (for low metallicity and rapid rotation models) mass.
The yields of heavy element produced by CCSN ejecta also depend on the remnant mass of massive mass which is mainly determined
by the mass of CO-core. Our models calculate that the yields of heavy elements produced by CCSN ejecta can get up to several $M_\odot$.
Compared with stellar wind, CCSN ejecta has a greater contribution to the heavy elements in ISM.
We also compare the $^{56}$Ni yields by calculated in this work with observational estimate.
Our models only explain the $^{56}$Ni masses produced by faint SNe or normal SNe with
progenitor mass lower than about 25 $M_\odot$, and greatly underestimate the $^{56}$Ni masses produced by
stars with masses higher than about 30 $M_\odot$.
}

   \authorrunning{R.-Q. Wu et al. }            %author_head in even pages
   \titlerunning{Heavy elements contribution of massive stars
   }  % title_head in odd pages
   \maketitle

%________________________________________________ sections below
%
\section{Introduction}           %% first-level sections will be auto-capitalized
\label{section1}

Interstellar medium (ISM) is defined as follows:
atomic, gas ions, dust grains, cosmic rays, and also includes many molecules.
Heavy elements are fundamental components to ISM and play a critical role in the stellar evolution
of astrophysics and the chemical evolution in the ISM.
It is well known that massive stars with a initial mass larger than $\sim 8 M_{\odot}$
play the most important role for producing the heavy elements in ISM \citep[e. g.,][]{Dunne2003,Ablimit2018,Du2020}.
These massive stars contribute to heavy elements via stellar wind and the ejecta of core$-$collapse
supernovae (CCSN). Although the heavy elements may originate from other sources
including the stellar wind of asymptotic giant branch stars, ejecta of classical novae, binary merger et al,
their contribution is very low
\citep{Groenewegen1993,Marigo2007,Hix2001,Lu2013,Zhu2013,Jose2006,Li2016,Rukeya2017,Zhu2019,Duolikun2019,Shi2020,Guo2020}.

The yields of heavy elements from massive stars have been investigated by many literatures
\citep[e. g.,][]{Woosley1995,Chieffi2004,Nomoto2006,Heger2010,Nomoto2013}.
However, these works do not consider mass loss which is very important for the massive star evolution
\citep[A recent review can be seen in][]{Smith2014}.
Usually, the mass loss was thought to be caused by stellar wind driven
by drastic radiation \citep{Castor1975,Puls2008}.
Simultaneously, it is also affected by metallicity and rotation
\citep{Vink2000,Vink2001,Meynet2000,Maeder2012}.
Because the efficiency of radiation pressure removing the stellar envelope depends on metallicity,
it has an effect on the mass-loss rates ($\dot{M}$) of massive stars by $\dot{M}\propto Z^{m}$,
where the index range of $m$ is from 0.5 to 0.94 \citep{Vink2000,Vink2001,Mokiem2007}.
Rotation can enhance the mass-loss rate \citep{Langer1998,Heger1998}. More importantly,
rapidly rotation can result in quasi chemically homogeneous evolution (CHE) induced by various instability,
such as dynamical shear instability, the Solberg-Hi{\o}land instability, the secular shear instability,
the Eddington-Sweet circulation, and the Goldreich Schubert-Fricke instability
\citep[e. g.,][]{Pinsonneault1989,Heger2000}.
CHE can carry the heavy elements produced by nuclear burning in the core to the stellar surface,
thus these heavy elements enter ISM via stellar wind \citep[e. g.,][]{Brott2011,Song2016,Cui2018}.
The role of heavy elements mixing is critical,
it will affect the opacity of the envelope and increase the
luminosity and effective temperature of the star \citep{Glebbeek2009}.

The standard non-rotating single-star model is strongly opposed as a possible progenitor of the Supernovae (SNe)
\citep{Fremling2014,Bersten2014a}.
But \cite{Prantzos2018} recently gave the heavy element yields of rotating massive stars.
They considered effects of three initial rotational velocities, namely, 0, 150 and 300 km s$^{-1}$.
Initial velocity above $\sim 350$ km s$^{-1}$ more likely attain to the critical velocity \citep{Meynet2006}.
Furthermore, in the VLT-FLAMES Tarantula Survey,
\cite{Dufton2013} found that the projected rotational velocities of single early B-type stars
can reach approximately 450 km s$^{-1}$.
In binary systems, owing to mass transfer rotation velocity will reach to the Kepler velocity \citep{deMink2013}.

Meanwhile, rapidly rotation result in a more massive helium core via CHE
\citep{Belczynski2016,Eldridge2016,Mandel2016,Marchant2016,Wang2018}.
On account of the helium-core masses have greatly effects on the remnant masses of neutron stars
and black holes \citep[e. g.,][]{Hurley2000,Belczynski2008}.
At the pre-supernova (pre-SN) stage, a larger helium-core burning produces a bigger CO-core
\citep{Meynet2006,Kohler2015,Marassi2019}.
Hence, rotation as well as affects the heavy elements of CCSN ejecta.
Very recently, in order to study dust formation in CCSN ejecta,
\cite{Marassi2019} considered the effects of rotation, metallicity,
and fallback, they computed the heavy element yields of massive stars.
However, they did not still considered the yields via stellar wind.

Therefore, it is necessary to study the heavy element yields coming from stellar wind and CCSN ejeca for
massive stars. Even the research on the relevant factors of elemental abundance is very urgent.
In this paper, we study the effects of metallicity, rotation and fallback on the contribution
of heavy elements produced by massive stars.
In Section 2, the input physical parameters in models are described.
The detailed results are discussed in Section 3.
The main conclusions appear in Section 4.

\section{Model}\label{sec:2}

We use the open-source stellar evolution code Modules Experiments in Stellar Astrophysics
(MESA, version 10108, model core-collapsed supernova) to simulate massive star evolutions \citep{Paxton2011,Paxton2013,Paxton2015}.
In these simulations, we select 67$-$isotope network.
The mixing-length parameter ($\alpha_{\rm mlt}$) is taken as 1.5 \citep{Brott2011,Moravveji2016,Ma2020,Shi2020}.
In addition, the Ledoux criterion connects with boundaries of convection,
semi-convection ($\alpha_{\rm sc}$) is selected as 0.02.
Most of all, MESA has the Ledoux criterion $\nabla=\nabla_{\rm rad}$ in the overshoot area,
which is different with deep overshoot method \citep{Maeder1975,Viallet2015}.
Overshooting between convective core and radiative of interior
diffusion parameter is expressed by ($f_{\rm ov}=0.05$).
Another effective parameter ($f_{\rm o}=0.02$) is from the surface down to overshoot layer,
\citep{Paxton2011,Moravveji2016,Higgins2019}, they are considered at all stages of evolution,
they also can affect the total mass of stellar loss.
Thermohaline mixing parameter ($\alpha_{\rm th}$) is equal to $2.0$ \citep{Kippenhahn1980,Paxton2013}.
In this work, we use the formulae of \cite{Vink2001} to calculate the mass-loss rates.
In addition, rotation can enhance mass-loss rate by
\begin{equation}
\dot{M}(\Omega)=(\frac{1}{1-\Omega/\Omega_{\rm crit}})^\gamma{\dot{M}(0)},
\label{eq:1}
\end{equation}
where $\dot{M}(0)$ is the mass$-$loss rate without rotation,
$\Omega$ and $\Omega_{\rm crit}$ represent the angular velocity and critical Keplerian
angular velocity, respectively, and parameter $\gamma$ equals $0.43$ \citep{Langer1998}.
$\dot{M}(0)$ is calculated by the formulae in \cite{Vink2001}.
But when the angular velocity reaches the critical angular velocity, there will be a singularity.
We limit the mass loss rate so that the mass loss time scale is longer than the thermal time scale of the star,
see (1)-(3) equations in \cite{Yoon2012}.

In order to discuss the effects of metallicity,
the 4 initial metallicities are taken in different models as follows:
$[Fe/H]=0$, $[Fe/H]=-1$, $[Fe/H]=-2$ and $[Fe/H]=-3$.
Here, $[Fe/H]=\log [(Fe/H)/(Fe/H)_{\odot}]$ where $[Fe/H]_{\odot}=0.02$
is the solar metallicity \citep{Thielemann2010,chiaki2015}.

Considering that rotational velocity of massive stars
may get up to the critical velocity \citep{deMink2013} at the stellar surface,
we take the initial rotational velocities in different
simulations as 0, 300 and 500 km s$^{-1}$, respectively.
Rotation triggers some instabilities, then lead to angular
momentum transport and chemical mixing \citep[e. g.,][]{Meynet2012}.
Based on the research of \cite{Pinsonneault1989}, \cite{Heger2000} and \cite{Yoon2006},
MESA uses the ratio of the turbulent viscosity to the diffusion
coefficient ($f_{\rm c}$) and the ratio of sensitivity to
chemical gradients ($f_{\rm \mu}$) to calculate angular momentum transport
and chemical mixing induced by rotation.
\cite{Zhu2017} and \cite{Cui2018}
employed MESA to investigate rotating massive stars.
Following them, we choose $f_{\rm c}=$0.0228 and $f_{\rm \mu}=$0.1, respectively.

MESA code can calculate the stellar evolution from pre-main sequence to CCSN.
However, it can not give the remnant masses after CCSN.
In our work, following \cite{Hurley2000,Belczynski2008,Wang2018}
the remnant masses of NS or BH are given by CO-core mass.
CCSNe model of the MESA via collapsing of an core when the mass of Fe core $> 1.4$ M$_\odot$,
and we do not consider the nuclear reaction in this stage.
The explosion mechanism of CCSNe is a complex process which is still has not been well explained.
In our model the explosion energy (E) is $1\times10^{51}$ ergs
\citep{Nomoto2007,Paxton2013,Hirschi2017,Curtis2019}.

Simultaneously, a supernova explosion occurs when stellar central density gets to
$7.9 \times 10^{9}$ $g/cm^{3}$
and central temperature is $\sim6.5 \times 10^{9} k$.

%their nuclear time-scale is about $t_{\rm pre}\sim [1-15]\times10^{6} {\rm yr}^{-1}$ \citep{Witt2001}.
\begin{figure}[h!]%"[]"中为位置参数，四个参数tbph依次是置顶、置底、浮动、当前位置，，选用的参数优先顺序为h-t-b-p
\centering
\includegraphics[scale=0.3, angle=-90]{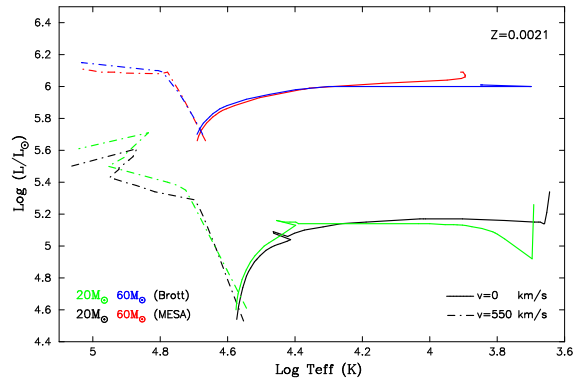}%"scale"后的数字为图形的宽度，也可用"width=1.0\columnwidth"定义
\caption{The evolutions of massive stars with different masses and rotation velocities for
$Z=0.0021$. The solid lines represent a non-rotating star,
while the dash-dotted lines represent a star with a rotation velocity of 550 km s$^{-1}$ .
Green and blue lines are the evolutional tracks calculated by \cite{Brott2011},
and black and red lines are simulated in our models.} %图题
\label{Fig1}%{}中"fig:example1"为图名，引用时用\ref{fig:example1}
\end{figure}

\section{Results}\label{sec:3}

Using MESA code, we simulate the evolutions from main sequence (MS) to CCSN for
8 massive stars with masses of 13, 15, 20, 25, 30, 40, 60 and 80 M$_\odot$.
In order to discuss the effects of rotation,
the initial rotational velocities are
taken 0, 300 and 500 km s$^{-1}$ in different simulations, respectively.
In order to check our model,
we compare the evolutions of several stars with
these in \cite{Brott2011} under similar input parameters.
Figure~\ref{Fig1} shows the evolutional tracks in two works are similar.
All heavy elements originating from star are produced by
nucleosynthesis. They are ejected into ISM via stellar wind and CCSN ejecta.

\subsection{Heavy elements coming from stellar wind}\label{sec:3.1}

Before massive stars occur CCSN, their heavy elements enter ISM via stellar winds.
These heavy elements locate in stellar envelope.
In this work, we estimate the yields of i-th heavy element by
\begin{equation}
M_{\rm i}=\int_{\rm 0}^{t_{\rm pre}}\dot{M}(t)X_{\rm i}(t){\rm d}t,
\label{yields}
\end{equation}
where $t_{\rm pre}$ is the time from zero-age MS to pre-CCSN,
$\dot{M}(t)$ and $X_{\rm i}(t)$ are the mass-loss rate and the mass fraction of
i-th heavy element on the surface of massive star, respectively.
Therefore, the heavy elements coming from stellar wind mainly depend on the mass-loss
rates and the chemical abundances on the stellar surface.

\begin{figure}[h!]%"[]"中为位置参数，四个参数tbph依次是置顶、置底、浮动、当前位置，，选用的参数优先顺序为h-t-b-p
\centering
\includegraphics[scale=0.4
, angle=-90]{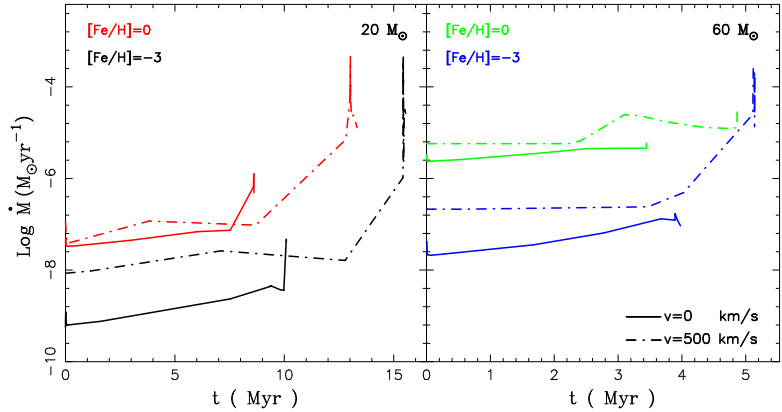}%"scale"后的数字为图形的宽度，也可用"width=1.0\columnwidth"定义
\caption{The evolutions of mass-loss rates for models with different masses (20 and 60 $M_\odot$),
 metallicities ($[Fe/H]=0, -3$) and rotational velocities (0 and 500 km s$^{-1}$).} %图题
\label{Fig2}%{}中"fig:example1"为图名，引用时用\ref{fig:example1}
\end{figure}

In our model, the mass-loss rates are affected by metallicity and rotational velocity.
Figure~\ref{Fig2} shows the evolutions of mass-loss rates for different initial mass
stars with different metallicities and rotational velocities.

\begin{figure}[h!]%"[]"中为位置参数，四个参数tbph依次是置顶、置底、浮动、当前位置，，选用的参数优先顺序为h-t-b-p
\centering
\includegraphics[scale=0.35, angle=-90]{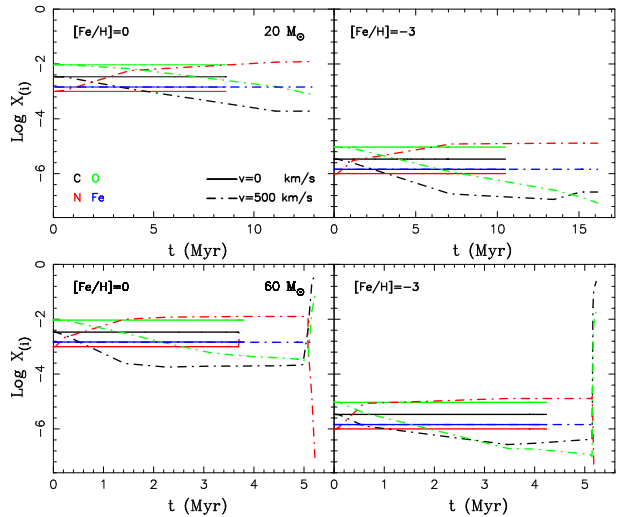}%"scale"后的数字为图形的宽度，也可用"width=1.0\columnwidth"定义
\caption{The evolutions of heavy-element abundances
[$^{12}$C (black), $^{14}$N (red), $^{16}$O (green), and $^{56}$Fe (blue)] on the stellar surfaces for massive stars.
The two panels in the top region represent the models with 20 $M_\odot$, while the two panels in the
bottom region is for 60 $M_\odot$. The solid and dash-dotted lines represent models with $V=0$ and 500 km s$^{-1}$,
respectively. The different metallicities are given in the left-top region of every panel.} %图题
\label{Fig3}%{}中"fig:example1"为图名，引用时用\ref{fig:example1}
\end{figure}

Compared the two metallicities models,
a high metallicity can result in a high mass-loss rate because $\dot{M}\propto Z^{m}$, where parameter
$m$ ranges from 0.64 to 0.85 \citep{Vink2001}.
Simultaneously, the mass-loss rates depends on the rotational velocities by Eq.~\ref{eq:1} mainly for MS stage,
we also consider the red supergiant (RSG) or Wolf-Rayet (WR) stage \citep{Nugis2000}.
Therefore, the higher the initial rotational velocity is, the higher the mass-loss rates is.
The mass-loss rate can be enhanced about 1-4 magnitude when the initial rotational velocity increases
from 0 to 500 km s$^{-1}$.
The chemical abundances on stellar surfaces are determined by CHE.
During MS late phase, the star begins to rapidly expand, the rotational velocity sharply
decrease. Therefore, CHE mainly works in MS phase. The heavy elements affected by
nucleosynthesis during MS phase are $^{12}$C, $^{14}$N and $^{16}$O (key elements of the evolution of massive stars).

Figure~\ref{Fig3} gives the evolutions of heavy-element ($^{12}$C, $^{14}$N and $^{16}$O) abundances on the stellar surfaces.
Obviously, if there is no CHE in models without rotation,
the heavy-element abundances on the stellar surface are constant during its life.
However, in rotational models, the abundances of $^{12}$C, $^{14}$N and $^{16}$O elements on the stellar surfaces change.
$^{12}$C and $^{16}$O abundance decreases, while $^{14}$N abundances increases.
In particular, the lower metallicity is, the stronger CHE is. Therefore,
for lower metallicity models, the range of increase and decrease in abundance is more obvious.
In addition, for 60 $M_{\odot}$ panel,
because the H-rich shell are stripped out before the RSG phase, and the star come in a WR stage.
As Figure~\ref{Fig3} shows, $^{12}$C and $^{16}$O elements under stellar surface deeply increase while $^{14}$N elements decreases.
Similar results have been discussed in \cite{Maeder2001}, \cite{Hirschi2005a}, \cite{Chieffi2013}, \cite{Groh2014}, and \cite{Meyer2020}.

\begin{figure}[h!]%"[]"中为位置参数，四个参数tbph依次是置顶、置底、浮动、当前位置，，选用的参数优先顺序为h-t-b-p
\centering
\includegraphics[scale=0.35, angle=-90]{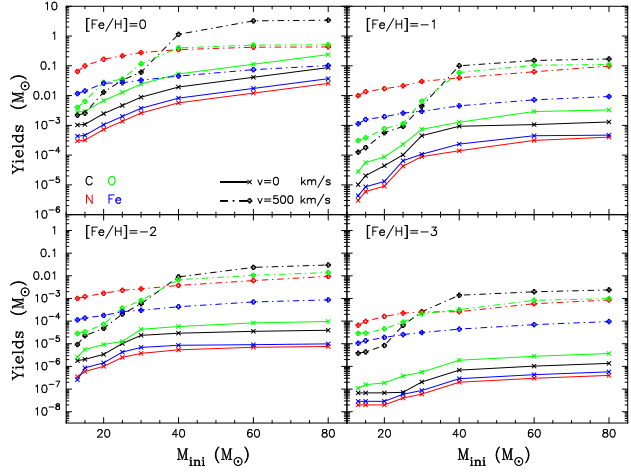}%"scale"后的数字为图形的宽度，也可用"width=1.0\columnwidth"定义
\caption{The yields of the heavy elements,
$^{12}$C (black), $^{14}$N (red), $^{16}$O (green), and $^{56}$Fe (blue),
produced via stellar winds from massive stars with different initial masses,
metallicities ($[Fe/H]=0, -1, -2, -3$),
and $V=$0, 500 km s$^{-1}$. The multiplication and addition symbols
represent calculated models with $V=0$ and 500 km s$^{-1}$, respectively.} %图题
\label{Fig4}%{}中"fig:example1"为图名，引用时用\ref{fig:example1}
\end{figure}
Figure~\ref{Fig4} gives the yields of heavy elements
($^{12}$C, $^{14}$N, $^{16}$O, and $^{56}$Fe) produced via stellar winds. \cite{Hirschi2005b} also calculated the
yields of heavy elements produced by stellar winds. In the Table 3 for a model
with $M_{\rm ini}=20 M_{\odot}$, $[Fe/H]=0$ and $V=300$ km s$^{-1}$,
\cite{Hirschi2005b} gave the yields of $^{12}$C, $^{14}$N, $^{16}$O elements are $1.73 \times10^{-2},
4.30 \times10^{-2}$ and $2.75 \times10^{-2} M_{\odot}$, respectively.
Under similar input parameters, the yields in our models are $3.34 \times10^{-2}, 7.20 \times10^{-2}$,
and $4.32 \times10^{-2} M_{\odot}$, respectively.
For a model with $M_{\rm ini}=40 M_{\odot}$, they in \cite{Hirschi2005b}
are $1.60, 1.73 \times10^{-1}$ and $3.34 \times10^{-1} M_{\odot}$, respectively.
They in our work are $1.15, 1.74 \times10^{-1}$ and $3.13 \times10^{-1} M_{\odot}$, respectively.
The results in both works are consistent.

In short, the yields of heavy elements coming from stellar winds can get to several $M_\odot$ for
high rotation and high metallicity, while they may only be $10^{-2}$ $M_\odot$ for low rotation
and low metallicity.

\begin{figure*}[htbp]%"[]"中为位置参数，四个参数tbph依次是置顶、置底、浮动、当前位置，，选用的参数优先顺序为h-t-b-p
\centering
\begin{tabular}{lr}
\includegraphics[scale=0.28, angle=-90]{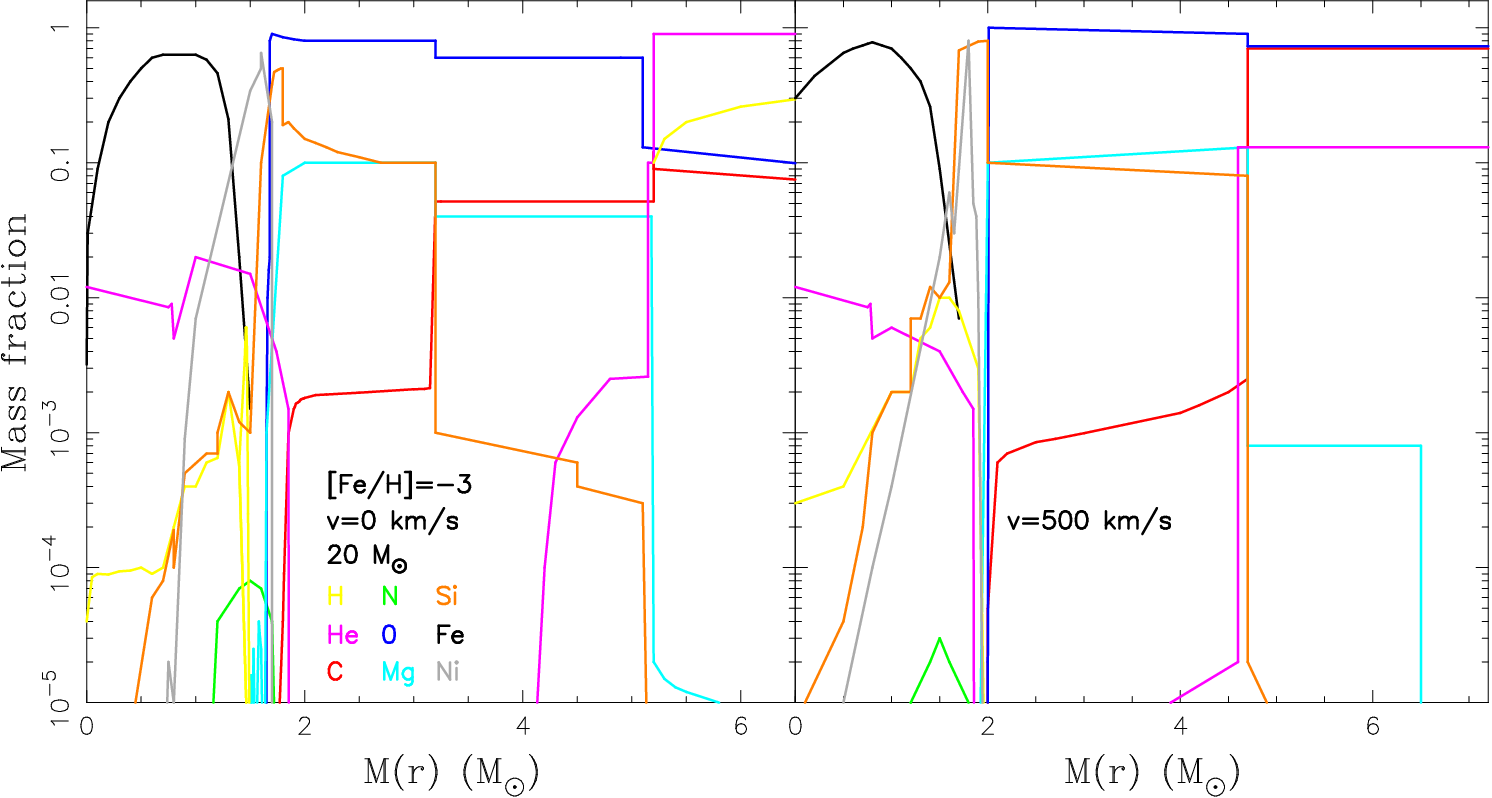}
\includegraphics[scale=0.28, angle=-90]{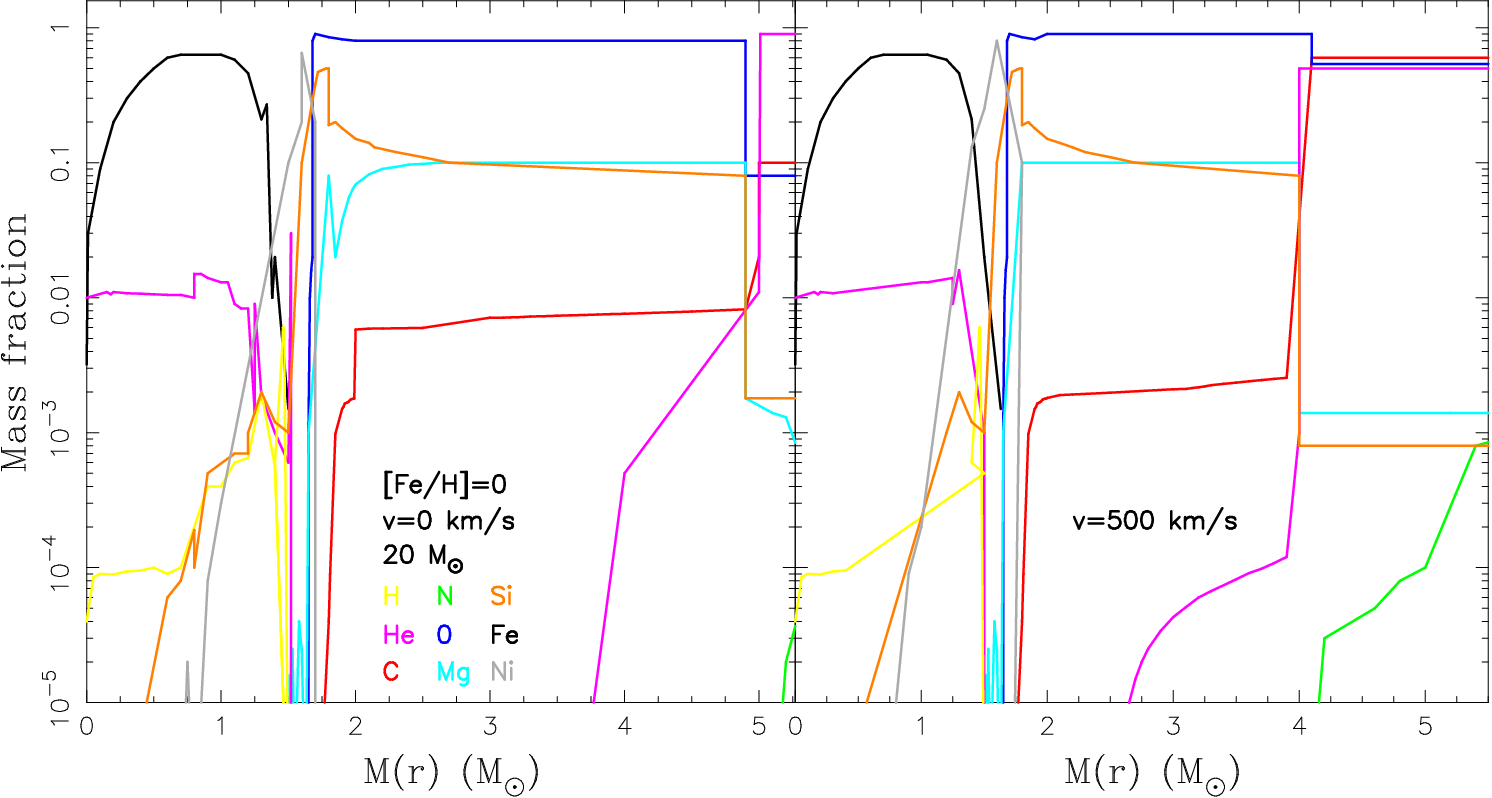}%"scale" 后的数字为图形的宽度，也可用"width=1.0\columnwidth"定义
\end{tabular}
\caption{The mass fractions of chemical elements in stellar interiors [M(r)] at pre-CCSN for
models with initial mass of 20 $M_\odot$.
The left two panels (the initial rotational velocities are 0 and 500 km s$^{-1}$, respectively)
represent the models with $[Fe/H]=-3$, and the right two panels for the models with $[Fe/H]=0$.
The abundance of various chemical elements is represented by colorful lines,
for example, $^{1}$H (yellow), $^{12}$C (red), $^{14}$N (green) etc.} %图题
\label{Fig5}%{}中"fig:example1"为图名，引用时用\ref{fig:example1}
\end{figure*}

\begin{figure*}[htbp]%"[]"中为位置参数，四个参数tbph依次是置顶、置底、浮动、当前位置，，选用的参数优先顺序为h-t-b-p
\centering
\begin{tabular}{lr}
\includegraphics[scale=0.28, angle=-90]{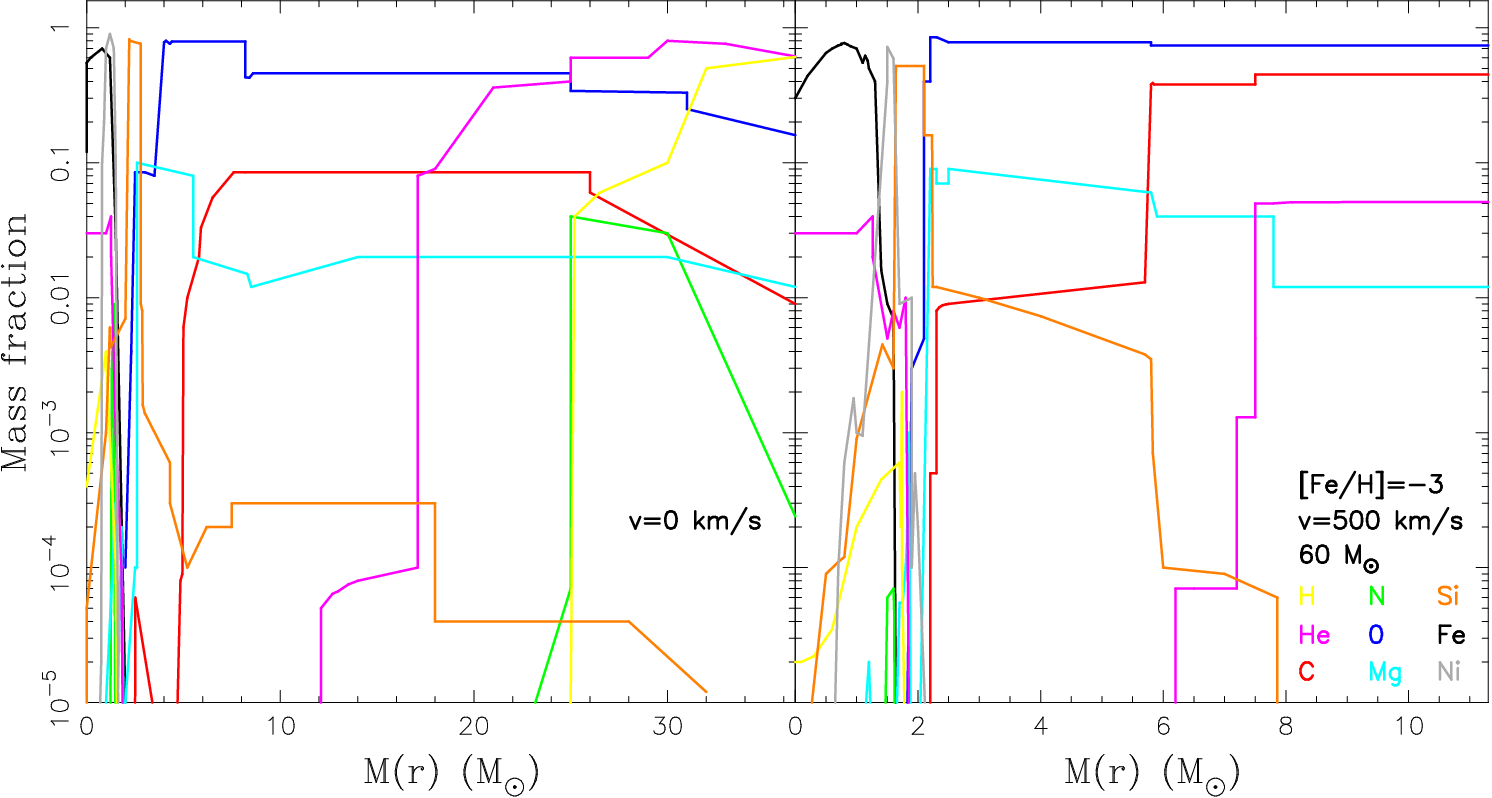}
\includegraphics[scale=0.28, angle=-90]{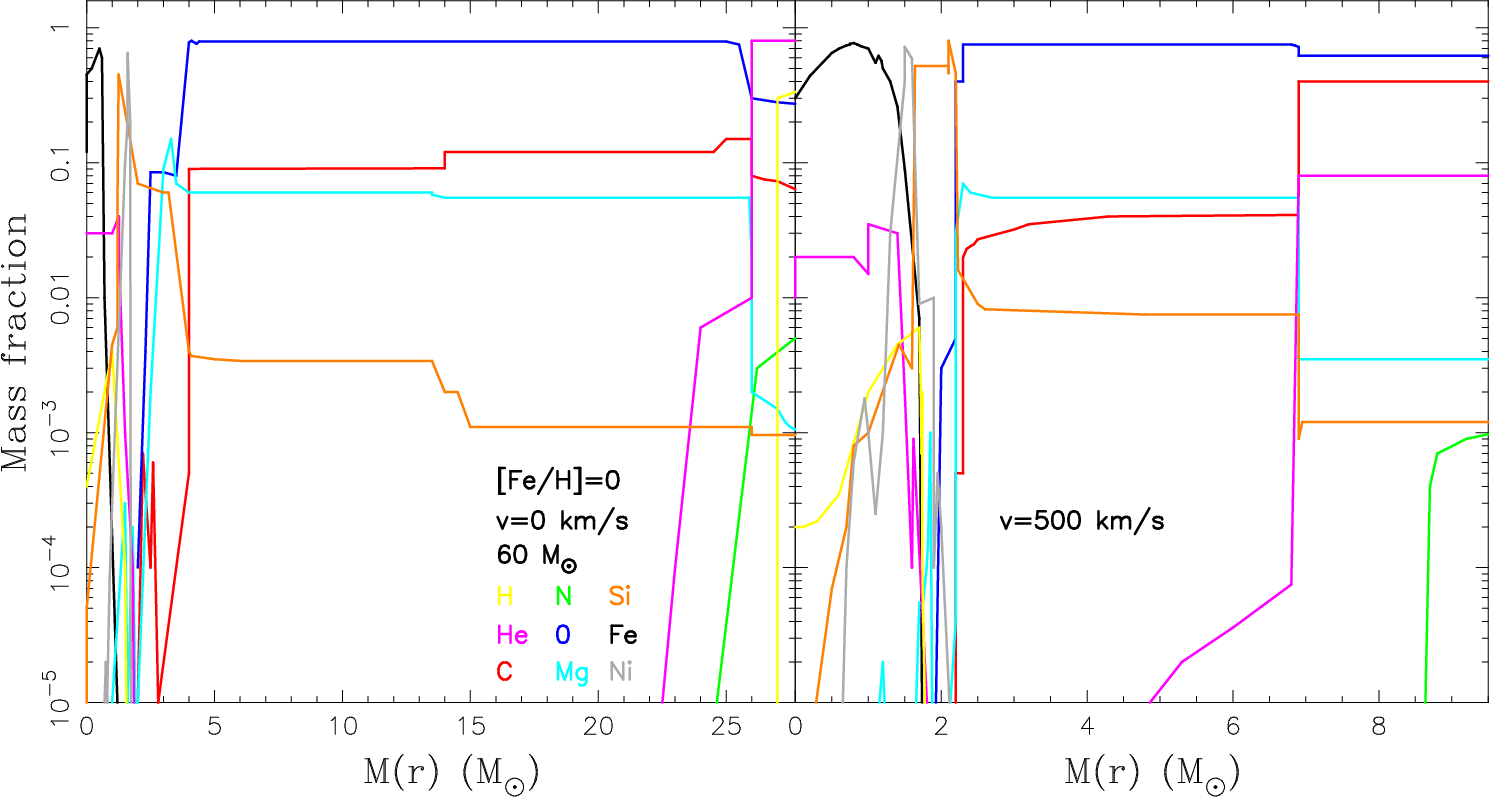}%"scale" 后的数字为图形的宽度，也可用"width=1.0\columnwidth"定义
\end{tabular}
\caption{Similar with Figure~\ref{Fig5} but for models with initial mass of 60 $M_\odot$.} %图题
\label{Fig6}%{}中"fig:example1"为图名，引用时用\ref{fig:example1}
\end{figure*}

\subsection{Heavy elements coming from SN ejecta}\label{sec:3.2}

The heavy elements locating in stellar interiors are ejected
into ISM via CCSN. They are mainly determined by mass fractions
before CCSN occurs.
Figure~\ref{Fig5} and Figure~\ref{Fig6} show the fractions of different elements in the models.
For models with a mass of 20 $M_\odot$, rapid rotation can enhance
mass-loss rates. The star with $V=500$ km s$^{-1}$ have lost
whole hydrogen envelope. Simultaneously, it can trigger CHE, produce a larger CO-core.
Therefore, compared with the star without rotation,
it has more massive core before CCSN.
The stars with low metallicity can undergoes efficient CHE and have
low mass-loss rate. Their CO-core at pre-CCSN are larger
than those for the stars with high metallicity.
Similar results appear in models with 60 $M_\odot$ star.
These results are consistent with these in \cite{vanMarle2007,Tominaga2008,Limongi2018}.
During CCSN, massive stars eject a portion of their masses and leave compact objects (neutron stars or black holes).
Generally, the remnant mass ($M_{\rm rem}$) is calculated by CO-core mass ($M_{\rm CO}$) \citep[e. g.,][]{Belczynski2008}.
In this work, we use the Equations (1) to (4) in \cite{Belczynski2008} to
calculate $M_{\rm rem}$.

\begin{figure*}[htbp]%"[]"中为位置参数，四个参数tbph依次是置顶、置底、浮动、当前位置，，选用的参数优先顺序为h-t-b-p
\centering
\begin{tabular}{lr}
\includegraphics[scale=0.28, angle=-90]{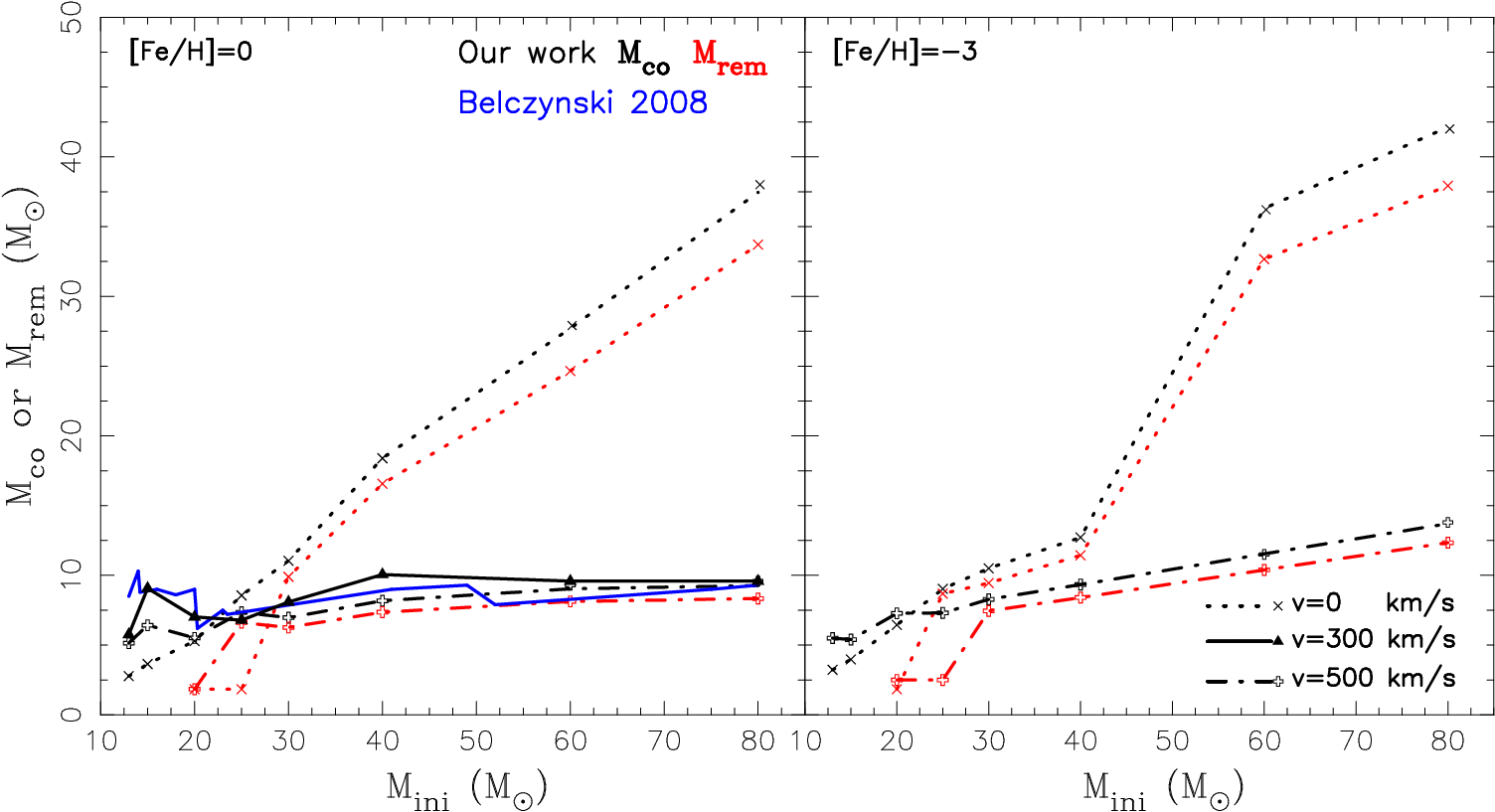}
\includegraphics[scale=0.28, angle=-90]{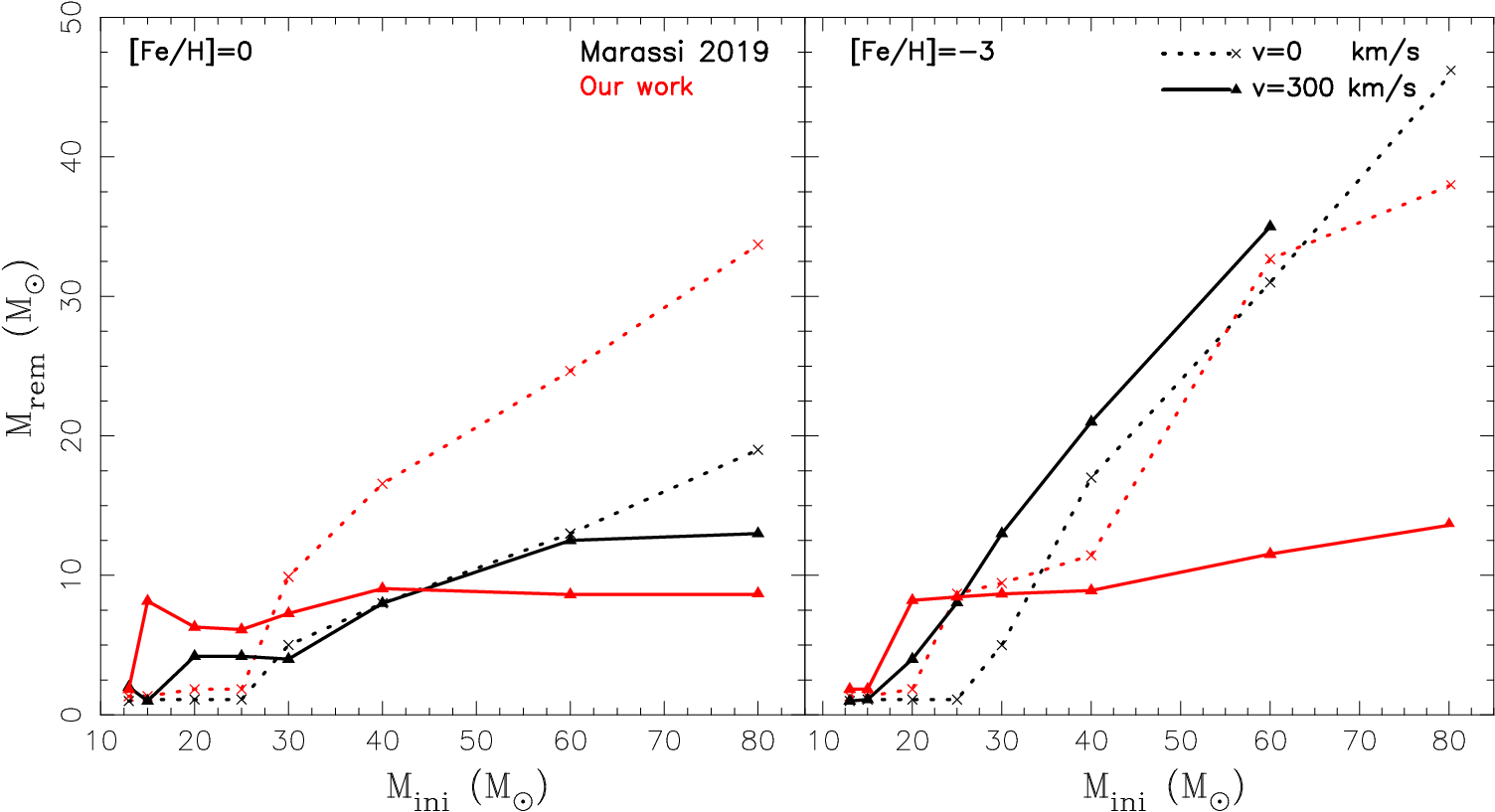}%"scale" 后的数字为图形的宽度，也可用"width=1.0\columnwidth"定义
\end{tabular}
\caption{Left: the CO-core ($M_{\rm CO}$) and remnant ($M_{\rm rem}$) masses vs. The initial masses in different models.
The black and the red lines represent $M_{\rm CO}$ and $M_{\rm rem}$, respectively (Our work).
The solid blue line represents the CO-core size of the \cite{Belczynski2008} model with a rotation of 300 km s$^{-1}$.
Right: comparison of the remnant masses in our work with those in \cite{Marassi2019}.
Black and red lines represent \cite{Marassi2019} results and ones in our.} %图题
\label{Fig7}%{}中"fig:example1"为图名，引用时用\ref{fig:example1}
\end{figure*}

Figure~\ref{Fig7} (Left) gives $M_{\rm CO}$ and $M_{\rm rem}$ calculated by different models.
$M_{\rm CO}$ and $M_{\rm rem}$ are mainly determined by mass-loss rates.
The stars with high metallicity and high rotational velocity have high mass-loss rates,
their $M_{\rm CO}$ and $M_{\rm rem}$ hardly exceed 10 $M_\odot$.
CHE triggered by rapid rotation can only increase the $M_{\rm CO}$ and $M_{\rm rem}$
of models with initial masses lower than about 30 $M_\odot$.
We compare the CO-core with the work of \cite{Belczynski2008} model with a rotation of 300 km s$^{-1}$,
obviously the size of the CO-core of the two models are consistent.

Figure~\ref{Fig7} (Right) compares $M_{\rm rem}$ calculated by this work with
those in \cite{Marassi2019}. Obviously, $M_{\rm rem}$ of stars with initial masses lower than
about 30 $M_\odot$ in our work is higher than that in \cite{Marassi2019},
while others in our work are lower. The main reasons are mass-loss rates and the method for calculating remnant mass.
For the former, as Figure~\ref{Fig2} in \cite{Marassi2019} showed, the hydrogen envelope with a mass of about 3 $M_\odot$ in
the model with 60 $M_\odot$ initial mass and [Fe/H]=-1 is left before CCSN. However, under the model with [Fe/H]=-1,
there is no hydrogen envelope for 60 $M_\odot$ initial mass at pre-CCSN.
For the latter, the remnant mass in \cite{Marassi2019} is determined by the initial mass and the metellicity,
while this work calculates $M_{\rm rem}$ via $M_{\rm CO}$.
Observation show that the CO nucleus will only appear when the gas density of the star reaches the standard value \cite{Chen2006}.
The yield of the i-th element produced by CCSN ejecta can be calculated by explosion.

\begin{figure*}[htbp]%"[]"中为位置参数，四个参数tbph依次是置顶、置底、浮动、当前位置，，选用的参数优先顺序为h-t-b-p
\centering
\begin{tabular}{lr}
\includegraphics[scale=0.3, angle=-90]{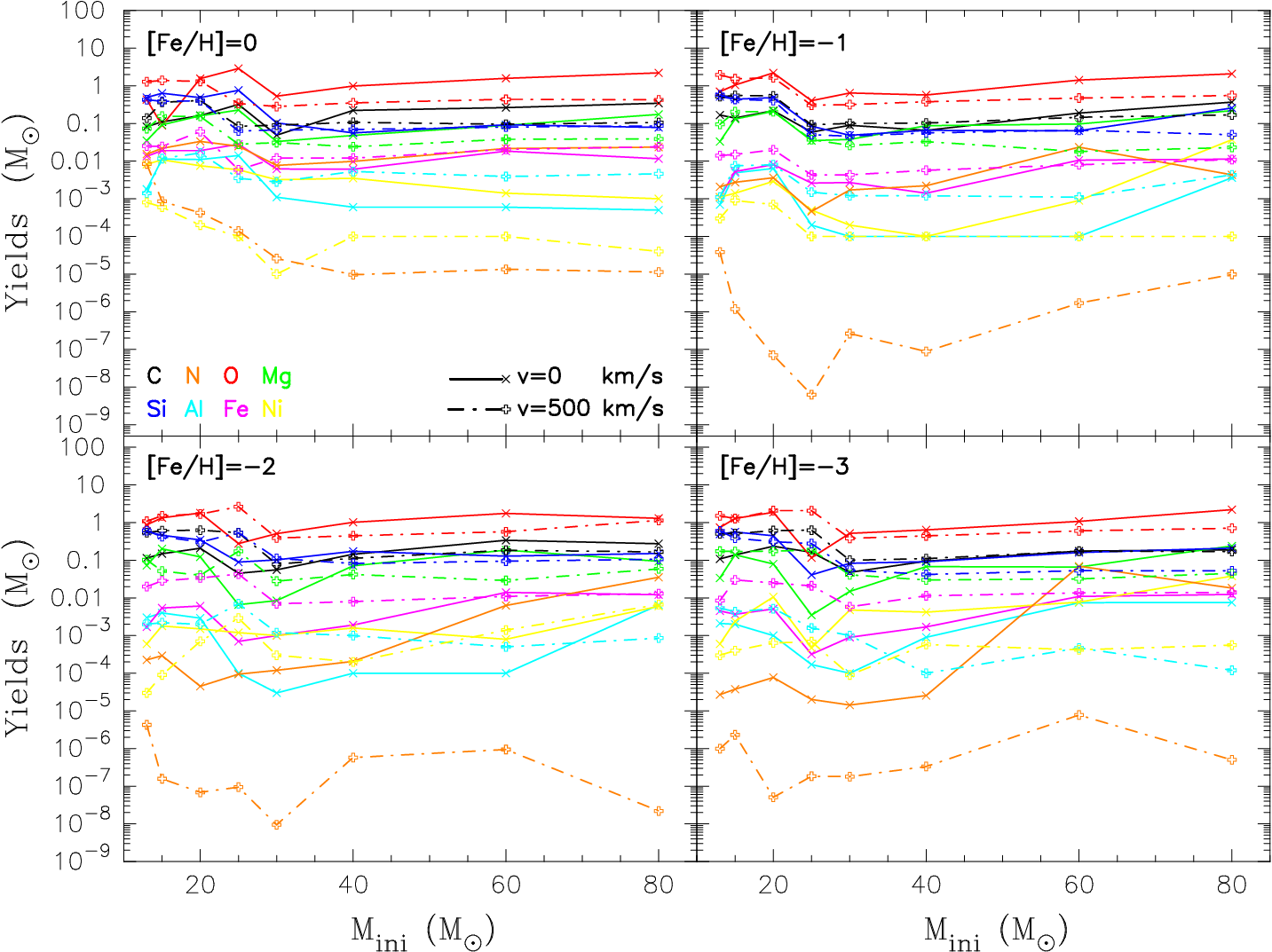}%"scale"后的数字为图形的宽度，也可用"width=1.0\columnwidth"定义
\includegraphics[scale=0.3, angle=-90]{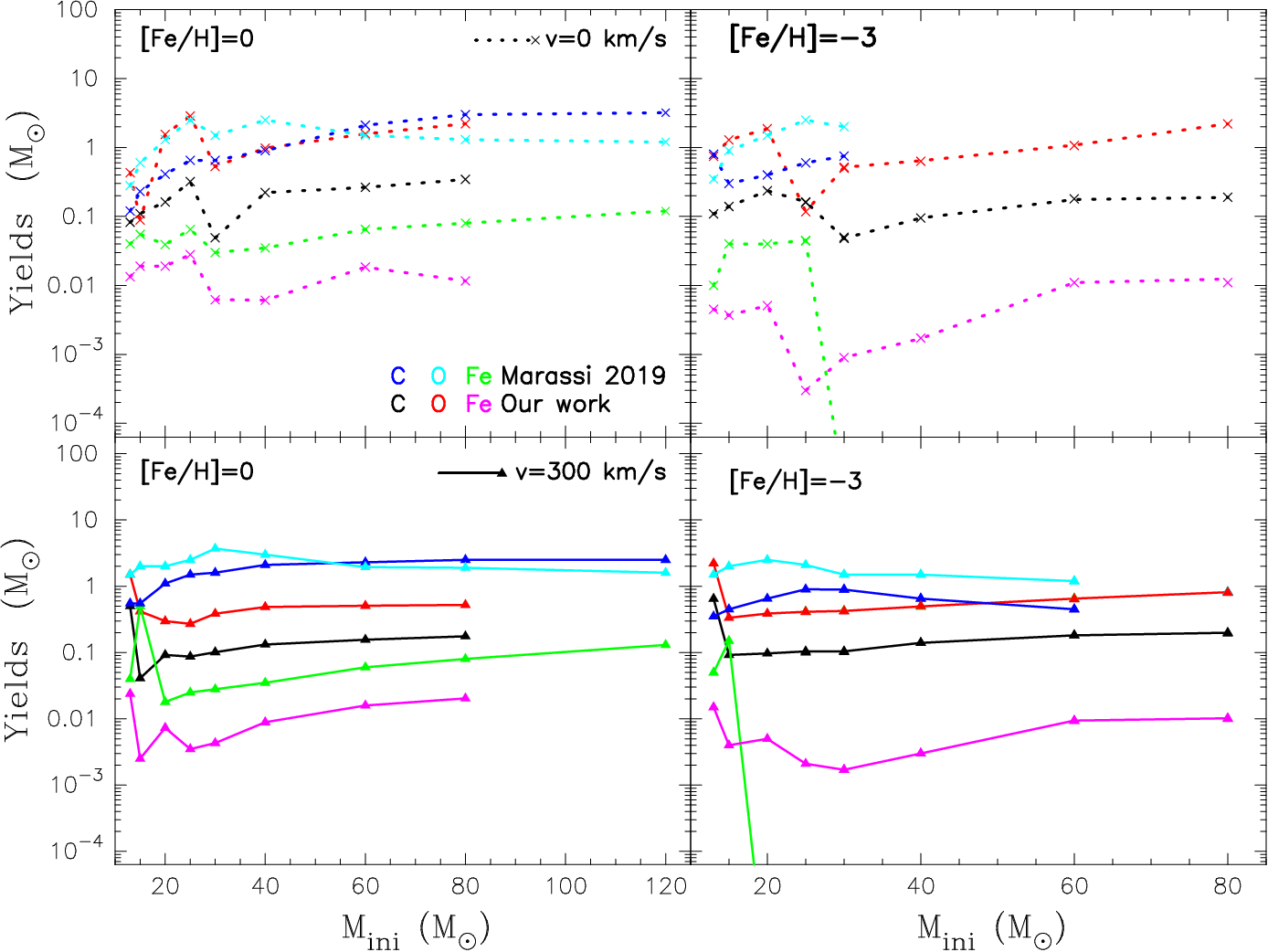}
\end{tabular}
\caption{Yields of heavy elements produced by CCSN ejecta. The left panel represents the results in
this work. The right panel gives the comparison of our results with ones in \cite{Marassi2019}.} %图题
\label{Fig8}%{}中"fig:example1"为图名，引用时用\ref{fig:example1}
\end{figure*}

\begin{figure}[h!]%"[]"中为位置参数，四个参数tbph依次是置顶、置底、浮动、当前位置，，选用的参数优先顺序为h-t-b-p
\centering
\includegraphics[scale=0.3, angle=-90]{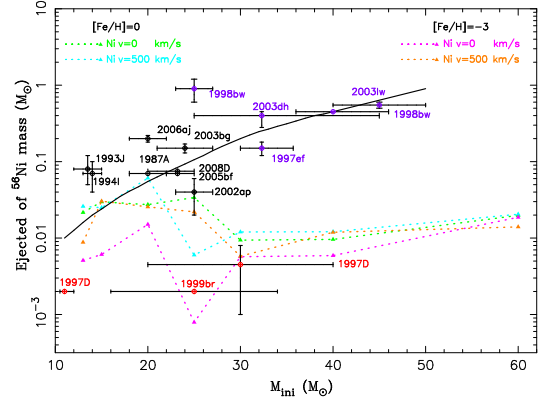}%"scale"后的数字为图形的宽度，也可用"width=1.0\columnwidth"定义
\caption{The $^{56}$Ni masses produced by CCSN ejecta and their progenitor
masses at MS phase. These data come from \cite{Nomoto2013}.
The red, black and purple cycles represent faint SNe, normal SNe and hypernovae,
respectively. The yields of $^{56}$Ni calculated by
our models are given by dotted lines ($[Fe/H]=0$) and dashed lines ($[Fe/H]=-3$).
For example, the green and blue colors represent models with $V=0$ and 500 km s$^{-1}$, respectively.} %图题
\label{Fig9}%{}中"fig:example1"为图名，引用时用\ref{fig:example1}
\end{figure}

Figure~\ref{Fig8} shows the yields of heavy elements produced by CCSN ejecta in this work.

\begin{equation}
M_{\rm i} = \int_{\rm M_{\rm rem}}^{M_{\rm fin}}[X_{\rm i}(m)-X_{\rm i}^{0}]{\rm d}m
\label{eq3}
\end{equation}
where $X_{\rm i}(m)$ is the i-th element mass fraction before the SN with Lagrangian coordinate m,
$X_{\rm i}^{0}$ is the initial abundance, the $X_{\rm i}^{0}$ of these elements are 0. $M_{\rm fin}$ is the final mass of the star \cite{Ekstrom2008}.
However, compared with stellar winds (see Figure~\ref{Fig4}),
CCSN ejecta can produce more heavy elements, especially the elements heavier than $^{16}$O element.
Compared with \cite{Marassi2019}, our work gives lower yields of heavy elements.
The main reason is that our models have higher mass-loss rates. Before CCSN occurs,
the stars in our models have lost more mass than those in \cite{Marassi2019}.

Via the comparison of observed light curves and the theoretical models, \cite{Nomoto2013}
estimated the $^{56}$Ni masses produced by some CCSN ejecta and their progenitor masses,
which are showed in Figure~\ref{Fig9}.
Here, $^{56}$Ni is caused by the decay of $^{56}$Ni $\rightarrow$ $^{56}$Co $\rightarrow$ $^{56}$Fe \citep{Argast2002,Hamuy2003}.
We calculate the yields of $^{56}$Ni in the different initial mass models.
Similar with the fixed energy models in \cite{Marassi2019},
our results only explain the $^{56}$Ni masses produced by faint SNe or normal SNe with
progenitor mass lower than 25 $M_\odot$.
Clearly, our understanding for the massive star evolution and CCSN is still poor.

\section{Conclusions}\label{sec:4}

In this work, we calculate the contribution of heavy elements from massive stars via stellar wind and
CCSN ejecta to ISM.

In our models, the evolutions of massive stars are affected
by rotation, mass-loss rate and metallicity.
The rotation via CHE changes the chemical abundances of stellar surfaces,
and enhances mass-loss rate. It can increase $^{14}$N abundance by 10 times while decrease
$^{12}$C and $^{16}$O abundances by similar times. It can enhance the mass-loss rates by about 1-4 magnitude
when the initial rotational velocity increases from 0 to 500 km s$^{-1}$.
Therefore, the yields of heavy elements coming from stellar winds are mainly affected by the stellar rotation.
We estimate that the stellar wind can produce heavy element yields of about $10^{-2}$ (for low metallicity models)
to several $M_\odot$ mass (for low metallicity and rapid rotaion models),
which depends on stellar rotation and metallicity.

The yields of heavy element produced by CCSN ejecta depend on not only rotation, mass-loss rate and metallicity,
but also the remnant mass of massive mass. Here, the latter mainly depends
on the mass of CO-core which is greatly affected by the above three parameters.
Our models calculate that the yields of heavy elements produced by CCSN ejecta can get up to several $M_\odot$ mass.
Compared with stellar wind, CCSN ejecta have a greater contribution to the heavy elements in ISM.

We also compare the $^{56}$Ni yields by calculated in this work with observational estimate.
Our models only explain the $^{56}$Ni masses produced by faint SNe or normal SNe with
progenitor mass lower than about 25 $M_\odot$, and greatly underestimate the $^{56}$Ni masses produced by
stars with initial masses higher than about 30 $M_\odot$.
It means that there is still a long way to understand the massive star and CCSN evolution.

\normalem
\begin{acknowledgements}
This work received the generous support of the National
Natural Science Foundation of China, projects No, 11763007, 11863005,
11803026, and U2031204.
We would also like to express our gratitude to the
Tianshan Youth Project of Xinjiang No. 2017Q014.

\end{acknowledgements}

\bibliographystyle{raa}
\bibliography{bibtex}

\end{document}